\documentclass{appolb}
\usepackage{graphicx}
\usepackage{subfigure}
\usepackage{setspace}
\usepackage{stackengine}

\usepackage{tikz-feynman}
\usepackage{pgfplots}
\usepgflibrary{plotmarks}
\usepackage{pgfplots}
\usepackage{xcolor,colortbl}
\usepackage{color, colortbl}
\definecolor{LightCyan}{rgb}{0.88,1,1}
\definecolor{darkcyan}{rgb}{0.,0.55,0.55}
\definecolor{darkgreen}{rgb}{0.1,0.6,0.1}
\definecolor{darkyellow}{rgb}{1,0.9,0.0}
\definecolor{metallicRed}{RGB}{135,22,20}
\definecolor{metallicBlue}{RGB}{44, 88, 128}
\definecolor{dkgreen}{rgb}{0,0.6,0}
\definecolor{gray}{rgb}{0.5,0.5,0.5}
\definecolor{mauve}{rgb}{0.58,0,0.82}

\begin{document}
\title{Light Pentaquark Searches with Hadron Beams%
}
\author{Jung Keun Ahn
\address{Department of Physics, Korea University, Seoul 02841, Republic of Korea}
}
\maketitle
\begin{abstract}
This review deals with measurements and future experiments of light pentaquark searches using hadron beams. 
\end{abstract}
  
\section{Introduction}

Pentaquarks are predicted to be made of four quarks and one antiquark,
forming the simplest multiquark configuration in the baryon sector. In contrast, tetraquark
states in the meson sector are firmly established.  
The discovery of numerous multiquark candidates 
poses questions about the dripline of multiquark states,  such as hexaquark states, in particular, 
the $H$-dibaryon.  
The observation of new multiquark states in different flavor sectors not only
tests the flavor-spin symmetry but may also lead to the discovery of new symmetries.

Light pentaquark states in the $KN$ system ($S=+1$) cannot be accommodated
in a conventional quark model for baryons ($3q$).
In the chiral quark soliton model,
the $\Theta^+$ is identified as a member of the $\overline{10}$
multiplet. The mass difference between the members of this multiplet was predicted
to be 180 MeV multiplied by their difference in strangeness unit. 
This prediction from the model highlights the low mass and narrow width of the $\Theta^+(1540)$, 
as well as its spin-parity of $J^p=1/2^+$, which is described as a rotating soliton \cite{dpp}. 

Early observations of the pentaquark $\Theta^+$ from photoproduction sparked 
a surge of pentaquark searches in various reactions. Initially, there were nearly equal numbers 
of positive and negative results from different experiments. 
However, over time, the weight of evidence began to shift against 
the existence of the $\Theta^+$ \cite{danilov, reflect, oset}. 
The story of the $\Theta^+$ came to a conclusion 
when the CLAS Collaboration found no evidence for its existence, 
even after analyzing data with statistics improved by two orders of magnitude.
The most stringent limit on the search for $\Theta^+$ is based on the CLAS results, which provides 
a 95\% C.L. upper limit on the cross section 0.7 nb for the 
$\gamma p\to\bar{K}^0K^+n$ and $\gamma p\to\bar{K}^0K^0p$  reactions \cite{clas1}.

\begin{figure}[!hbtp]
\centering
\subfigure[]{
\begin{tikzpicture}
\begin{feynman}
\vertex at (0, 1) (i1) {\(\gamma\)};
\vertex at (0,-1) (i2) {\(N\)};
\vertex at (1.5, 1) (a);
\vertex at (1.5,-1) (b);
\vertex at (3, 1) (f1) {\(\bar{K}\)};
\vertex at (3,-1) (f2) {\(\Theta^+\)};
\vertex at (2,0) () {\(K\)};
\diagram{(i1) -- [photon] (a),
(i2) -- [fermion] (b),
(f1) -- [anti fermion] (a),
(b) -- [fermion] (f2),
(a) -- [fermion] (b),};
\end{feynman}
\end{tikzpicture}
}
\subfigure[]{
\begin{tikzpicture}
\begin{feynman}
\vertex at (0, 1) (i1) {\(\gamma\)};
\vertex at (0,-1) (i2) {\(N\)};
\vertex at (1.5, 1) (a);
\vertex at (1.5,-1) (b);
\vertex at (3, 1) (f1) {\(K\)};
\vertex at (3,-1) (f2) {\(\Lambda(1520)\)};
\vertex at (2,0) () {\(K\)};
\diagram{(i1) -- [photon] (a),
(i2) -- [fermion] (b),
(f1) -- [anti fermion] (a),
(b) -- [fermion] (f2),
(b) -- [fermion] (a),};
\end{feynman}
\end{tikzpicture}
}
\subfigure[]{
\begin{tikzpicture}
\begin{feynman}
\vertex at (0, 1) (i1) {\(\gamma\)};
\vertex at (0,-1) (i2) {\(N\)};
\vertex at (1.5, 1) (a);
\vertex at (1.5,-1) (b);
\vertex at (3, 1) (f1) {\(\phi\)};
\vertex at (3,-1) (f2) {\(N\)};
\vertex at (2,0) () {\(P\)};
\diagram{(i1) -- [photon] (a),
(i2) -- [fermion] (b),
(f1) -- [anti fermion] (a),
(b) -- [fermion] (f2),
(b) -- [boson] (a),};
\end{feynman}
\end{tikzpicture}
}
\caption{(a) Feynman diagrams for $\gamma N\to \bar{K}KN$ reactions through 
production of either (a) $\Theta^+$, (b) $\Lambda(1520)$, or (c) $\phi$.
}
\label{fig:photoproduction}
\end{figure}
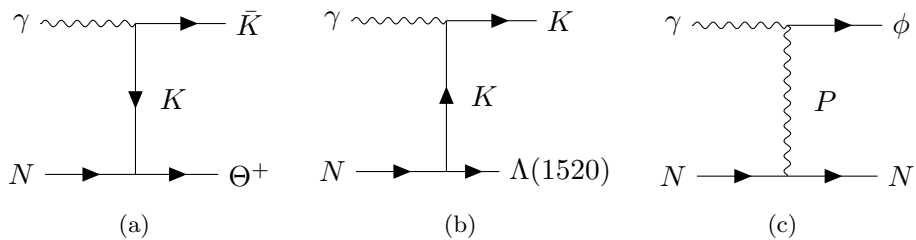

Despite this limit, there is still ongoing debate about how to conclude the $\Theta^+$ story, as 
high-statistics data have excluded its existence. The production of the $\Theta^+$ shares 
the same final state with either the $\phi$ or the $\Lambda(1520)$ in the $\gamma p$ and $\gamma d$ reactions, as depicted in Fig. \ref{fig:photoproduction}. Event selection for
$KN$ production at forward angles may enhance the visibility of the $\Theta^+$ signal in the presence
of significant background and the interference effects from 
the overlaping resonance bands \cite{amaryan2}. 
By imposing the condition $|t_\Theta|<0.45$ GeV$^2$, the $\Theta^+$ signal 
becomes distinctly noticeable within the same CLAS dataset \cite{amaryan}.  

The $\Theta^+$ therefore remains controversial, prompting us to consider 
what additional measurements could help clarify the situation. We anticipate
large production cross sections from hadron beams compared to those 
from electromagnetic processes. Pion-induced reactions yield different charge combinations 
in the final states due to the different charges of the intial states, which complement photoproduction proceses. These pion-induced and photoproduction reactions produce associated $\Theta^+$ particles when paired with $K^-$ or $\bar{K}^0$. On the other hand, kaon-induced
reactions provide a significant advantage by allowing for defining the strangeness. 

Therefore, the new data from $KN$ interactions is expected to considerably
impact the question about the existence of the $\Theta^+$. 
The $K^+N$ reactions involves two isospin states: $I=0$ and $I=1$. The isovector ($I=1$) states 
have been determined from elastic $K^+p$ reactions. Meanwhile, isoscalar states are determined 
from $K^+d$ breakup and elastic reactions, as well as from $K^0_L$ scattering off a proton. 
The reaction cross section in the $I=0$ state may provide interesting insight sine the dominant
$\Delta$ production mechanism is forbidden by isospin conservation, 
leaving other mechanisms more apparent \cite{hyslop}.
The potential existence of a narrow $\Theta^+$ state could manifest as a sudden change
in the inelastic cross section. 
This review will focus on the $K^+d$ and $K^+p$ reactions for the $\Theta^+$ study.

\section{$K^+p$ reactions}

\begin{figure}[htb]
\centerline{\stackinset{l}{2.6cm}{b}{1.9cm}{\small \textcolor{black}{$KN\pi$}}{
\stackinset{r}{1.2cm}{b}{1.0cm}{\small \textcolor{blue}{$KN\pi\pi$}}{
\stackinset{r}{0.9cm}{t}{2.0cm}{\small \textcolor{darkgreen}{$KN$ elastic}}{
\stackinset{l}{1.5cm}{t}{1.0cm}{\small \textcolor{red}{$KN$ total}}{
\includegraphics[width=0.475\textwidth]{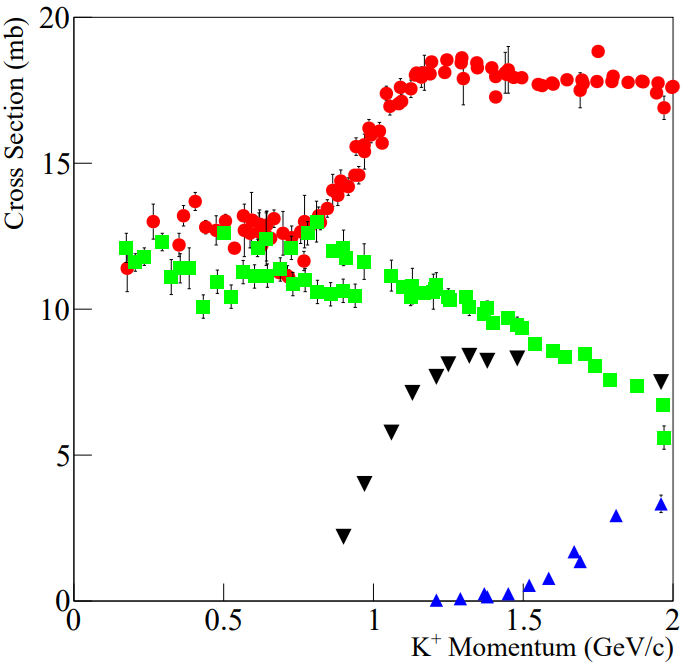} 
}}}}
\hskip+0.2cm
\stackinset{r}{1.0cm}{t}{1.0cm}{\small \textcolor{blue}{$K^+p\to pK^0\pi^+$}}{
\stackinset{r}{1.0cm}{t}{1.5cm}{\small \textcolor{darkgreen}{$K^+p\to pK^+\pi^0$}}{
\stackinset{r}{0.9cm}{t}{2.0cm}{\small \textcolor{red}{$K^+p\to nK^+\pi^+$}}{
\includegraphics[width=0.475\textwidth]{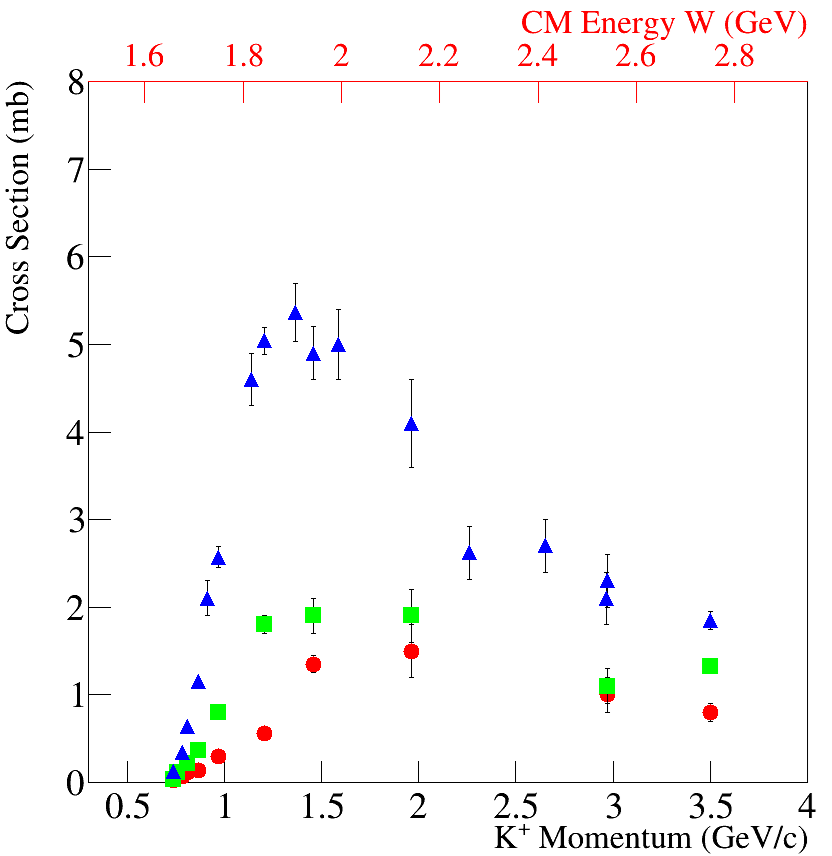} 
}}}
}
\caption{(a) $K^+p$ cross sections in the momentum range up to 2 GeV$/c$ and (b) $K^+p\to KN\pi$  cross sections in the momentum range up to 4 GeV$/c$ \cite{giacom}. }
\label{fig:KNpi}
\end{figure}

The $K^+p\to\Theta^+\pi^+$ two-body reaction is one of the most promising candidates 
for the search for the $\Theta^+$. In the $K^+p$ interaction, the elastic channel is dominant 
up to 0.8 GeV$/c$, and its cross section decreases gradually when the $KN\pi$ 
channels become accessible, as shown in Fig. \ref{fig:KNpi}(a). 
The cross sections for the $K^+p\to KN\pi$ reactions increase rapidly
near the threshold and reach the maximum at 1.5 GeV$/c$. In contrast, the cross section for 
the $KN2\pi$ channels increase slowly and remain quite small until 1.5 GeV$/c$.    
At around 1.5 GeV, the cross sections for the $K^+p$ elastic and $K^+p\to KN\pi$ channels 
are nearly equal. 

Old bubble chamber experiments provide cross section data for the 
$K^+p\to K^0p\pi^+$, $K^+p\to K^+p\pi^0$, and $K^+p\to K^+n\pi^+$ 
reactions between 1.2 and 1.7 GeV$/c$. 
There appears to be little nonresonant background, and the Dalitz plots 
indicate constructive interference at the crossing of the $\Delta$ and $K^\ast$ bands.
The production and decay of the $\Delta$ and $K^\ast$ resonances dominate
the $K^0p\pi^+$ final state.
Most features of the $KN$ spectra typically reflect these resonances \cite{berthon}. 
There is no clear indication of the $\Theta^+$ band. 
The $K^+p\to K^+n\pi^+$ reaction demonstrates a weak $\Delta$ resonance, 
although only 359 events have been recorded for this channel \cite{bland}. 

At approximately 1.5 GeV$/c$, two earlier bubble chamber experiments 
reported the cross sections for 
the $K^+p\to K^0p\pi^+$ reaction at momenta of 1.52 GeV$/c$ \cite{loken} 
and 1.585 GeV$/c$ \cite{bland}, where only the $K^\ast$ band intersects the phase space. 
The cross sections for the $KN\pi$ reactions are displayed for three different isospin channels 
in Fig. \ref{fig:KNpi}(b). The $\Theta^+$ was also searched for in the $K_Sp$ mass spectrum for
the $K^+p\to K_Sp\pi^+$  reaction at 11 GeV$/c$ \cite{lass}, but 
no evidence of the $\Theta^+$ signal was found.

\begin{figure}[!hbtp]
\centering
\subfigure[]{
\begin{tikzpicture}
\begin{feynman}
\vertex at (0, 1) (i1) {\(K^+\)};
\vertex at (0,-1) (i2) {\(p\)};
\vertex at (1.5, 1) (a);
\vertex at (1.5,-1) (b);
\vertex at (3, 1) (f1) {\(\pi^+\)};
\vertex at (3,-1) (f2) {\(\Theta^+\)};
\vertex at (2.0,0.) () {\(K^\ast\)};
\diagram{(i1) -- [fermion] (a),
(i2) -- [fermion] (b),
(f1) -- [anti fermion] (a),
(b) -- [fermion] (f2),
(a) -- [fermion] (b),};
\end{feynman}
\end{tikzpicture}
}
\subfigure[]{
\begin{tikzpicture}
\begin{feynman}
\vertex at (0, 1) (i1) {\(K^+\)};
\vertex at (0,-1) (i2) {\(p\)};
\vertex at (2.5,0) (a);
\vertex at (1,0) (b);
\vertex at (3.5, 1) (f1) {\(\pi^+\)};
\vertex at (3.5,-1) (f2) {\(\Theta^+\)};
\vertex at (1.75,-0.5) () {\(N\)};
\diagram{(i1) -- [fermion] (a),
(b) -- [anti fermion] (i2),
(f1) -- [anti fermion] (b),
(a) -- [fermion] (f2),
(a) -- [anti fermion] (b),};
\end{feynman}
\end{tikzpicture}
}
\caption{(a) Feynman diagrams for (a) the $t$-channel and (b) the $u$-channel processes 
in the $K^+p\to\Theta^+\pi^+$ reaction.
}
\label{fig:sigma}
\end{figure}
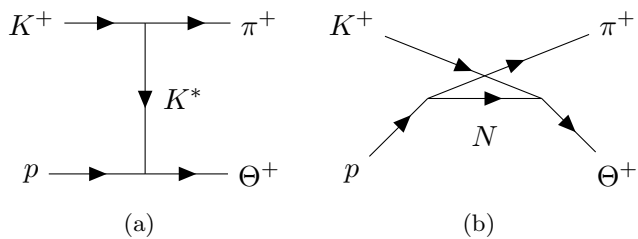

KEK-PS E559 searched for $\Theta^+$ via $(K^+,\pi^+)$ at a momentum of $1.2$ GeV$/c$ 
using a beam of $5\times 10^9$ $K^+$ \cite{miwa}. 
The analysis of the missing-mass spectrum for the $p(K^+,\pi^+)X$ reaction revealed 
no evidence of the $\Theta^+$ signal
within the forward lab angles ranging from $\theta^L_\pi=2^\circ$ to $22^\circ$.  
An upper limit of 3.5 $\mu$b$/$sr was established at a 90\% confidence level. 
These findings suggest either a suppression of the $K^\ast$ exchange contribution in the $t$ channel 
in Fig. \ref{fig:sigma}(a)
or points to a very small value of the coupling constant $g_{K^\ast N\Theta}$.   
If the $u$-channel processes in Fig. \ref{fig:sigma}(b) can only contribute to the $\Theta^+$ production, then the  $\pi^+$
angular distribution would be enhanced at backward angles. However, the E559 had no acceptance
in that angular region. 
Furthermore, the beam momentum $1.2$ GeV$/c$ was specifically chosen to maximize the cross sections for the $K^+p\to KN\pi^+$ reactions. However, the $\Delta^{++}$ and $K^\ast$ resonances
predominantly fill the phase space in the $K^0p\pi^+$ channel. 
At this momentum, the expected $\Theta^+$ band intersects with the two
resonance bands.

\begin{figure}[htb]
\centerline{%
\includegraphics[width=0.485\textwidth]{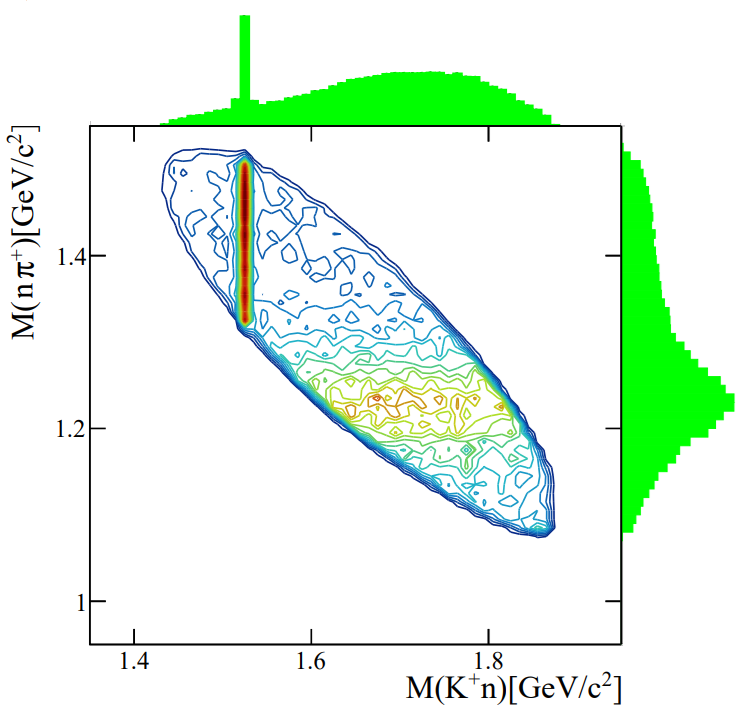} 
\hskip+0.2cm
\includegraphics[width=0.485\textwidth]{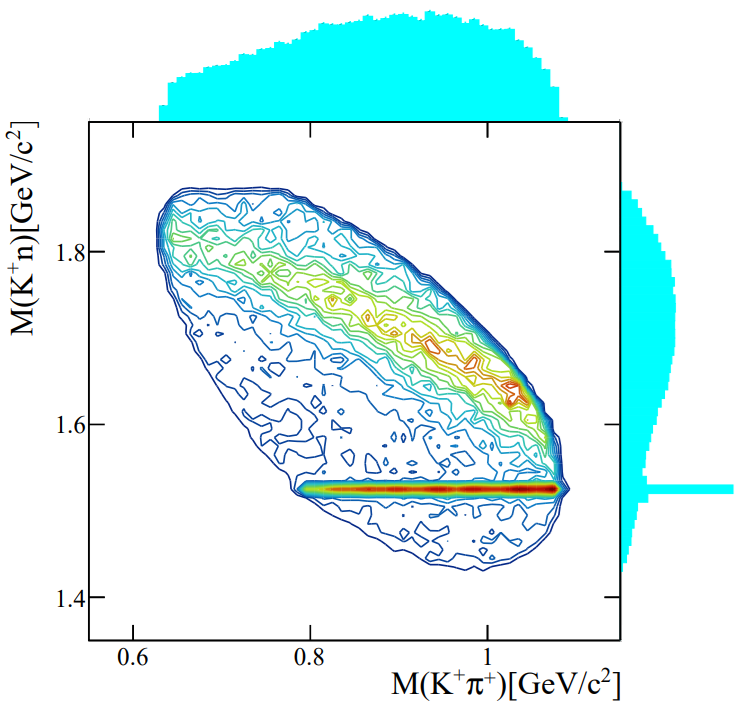} 
}
\caption{Dalitz plots of (a) the $K^+n$ vs $n\pi^+$ masses and (b) the $K^+\pi^+$ vs $K^+n$ masses 
for the $K^+p\to K^+n\pi^+$ reaction at $1.5$ GeV$/c$. }
\label{fig:kp}
\end{figure}

Therefore, a more definitive conclusion about the existence of the $\Theta^+$ will await 
further measurements. The $K^+p\to K^+n\pi^+$ reaction  
at a momentum of 1.5 GeV$/c$ is likely the most promising for the $\Theta^+$ search. 
This reaction does not involve any background from $K^\ast$ production, 
and only $\Delta$ resonance contributes to the available phase space. 
Moreover, the beam momentum of 1.5 GeV$/c$ ensures that 
the $\Delta$ resonance band does not overlap with the expected band for the $\Theta^+$, 
as illustrated in Fig. \ref{fig:kp}(a) and (b). The $\Theta^+$ peak appears superimposed 
on a smooth background shape in the simulated spectrum, which has a mass resolution of $3$ MeV. 
In contrast, in the $K^0p\pi^+$ channel, the $K^\ast$ band intersects 
the middle of the $\Theta^+$ band.

The experiment for the $K^+p\to K^+n\pi^+$ reaction  
can be conducted using a 1.5 GeV$/c$ $K^+$ beam in conjunction with
the Hyperon Spectrometer at J-PARC. 
The neutron can be reconstructed by detecting the outgoing $K^+$ 
and $\pi^+$ tracks using the HypTPC. The excellent separation of $\pi$ and $K$ 
provided by the HypTPC will help in identifying the $K^+p\to K^+n\pi^+$ reaction. 
Additionally, the HypTPC can also reconstruct the $K^0p\pi^+$ and $K^+p\pi^0$ channels.  

This experiment can provide crucial information on the spin and parity of
the $\Theta^+$. The conservation of parity in the $K^+p\to\Theta^+\pi^+$ reaction  
requires that the relative angular
momentum $L$ must be even. If the $\Theta^+$ has a spin of $1/2$, then $L$ must be zero.
Conversely, in the $K^+d\to\Theta^+p$ reaction, the opposite parity between the initial and final
states requires that the relative angular momentum $L$ be odd, and specifically $L=1$ 
for the $J=1/2^+ \Theta^+$.
The value of $L$ determines the angular distribution of the outgoing particle.
If the final state $\Theta^+\pi^+$ exhibits relative angular momentum zero,
the spin-parity of the $\Theta^+$ could also be $D5/2+$ ($L=2$) \cite{gibbs}. 
However, the production of the $\Theta^+$ competes with other strong background processes involving
$\Delta$ and $K^\ast$ resonances, making the unpolarized angular distributions insufficient to clarify the quantum numbers of the $\Theta^+$ \cite{hyodo}. Since the meson beams are unpolarized, utilizing target polarization can help determine the quantum number of the $\Theta^+$. 


\section{$K^+d$ reactions}

A direct formation of the $\Theta^+$ can occur through 
both the $K_Lp\to K^+n$ and $K^+n\to K^0p$ reactions. 
The $K_Lp\to K^+n$ reaction will be investigated using the KLF facility at Jefferson Lab \cite{klf}
and the details of this experiment will be presented in another review of this volume. 
The $K^+n\to K^0p$ reaction can also be realized through the $K^+d\to K^0pp$ reaction,
using a liquid deuterium target. 
Previous bubble chamber experiments have provided data with limited statistics
near the mass of the $\Theta^+$, although a few data points that deviate from a smooth trend have attracted attention.

\begin{figure}[htb]
\centerline{%
\stackinset{r}{1.6cm}{b}{1.0cm}{\small \textcolor{red}{$0^\circ$}}{
\stackinset{r}{1.0cm}{b}{1.9cm}{\small \textcolor{blue}{$10^\circ$}}{
\stackinset{r}{1.3cm}{b}{2.8cm}{\small \textcolor{darkcyan}{$20^\circ$}}{
\stackinset{l}{1.5cm}{b}{2.2cm}{\small \textcolor{black}{$K^+d\to\Theta^+p$}}{
\stackinset{r}{1.6cm}{t}{1.4cm}{\small \textcolor{red}{$0^\circ$}}{
\stackinset{r}{0.7cm}{t}{0.65cm}{\small \textcolor{blue}{$10^\circ$}}{
\stackinset{r}{1.3cm}{t}{0.3cm}{\small \textcolor{darkcyan}{$20^\circ$}}{
\stackinset{l}{1.2cm}{t}{0.5cm}{\small \textcolor{black}{$K^+p\to\Theta^+\pi^+$}}{
\includegraphics[width=0.485\textwidth]{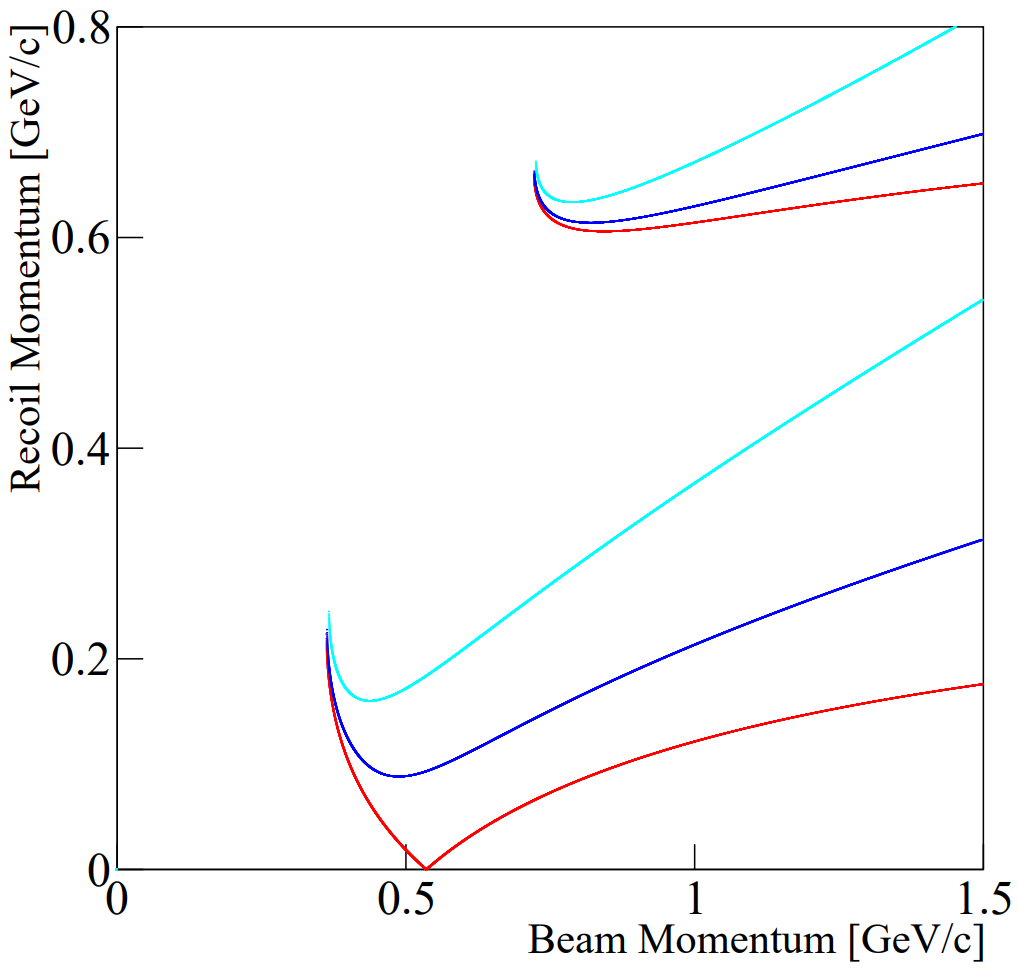} 
}}}}
}}}}
\hskip+0.2cm
\stackinset{r}{1.cm}{t}{1.5cm}{\small \textcolor{black}{$K^+d\to\Theta^+p$}}{
\stackinset{l}{1.65cm}{b}{2.2cm}{\small \textcolor{red}{$\Theta^+$}}{
\includegraphics[width=0.475\textwidth]{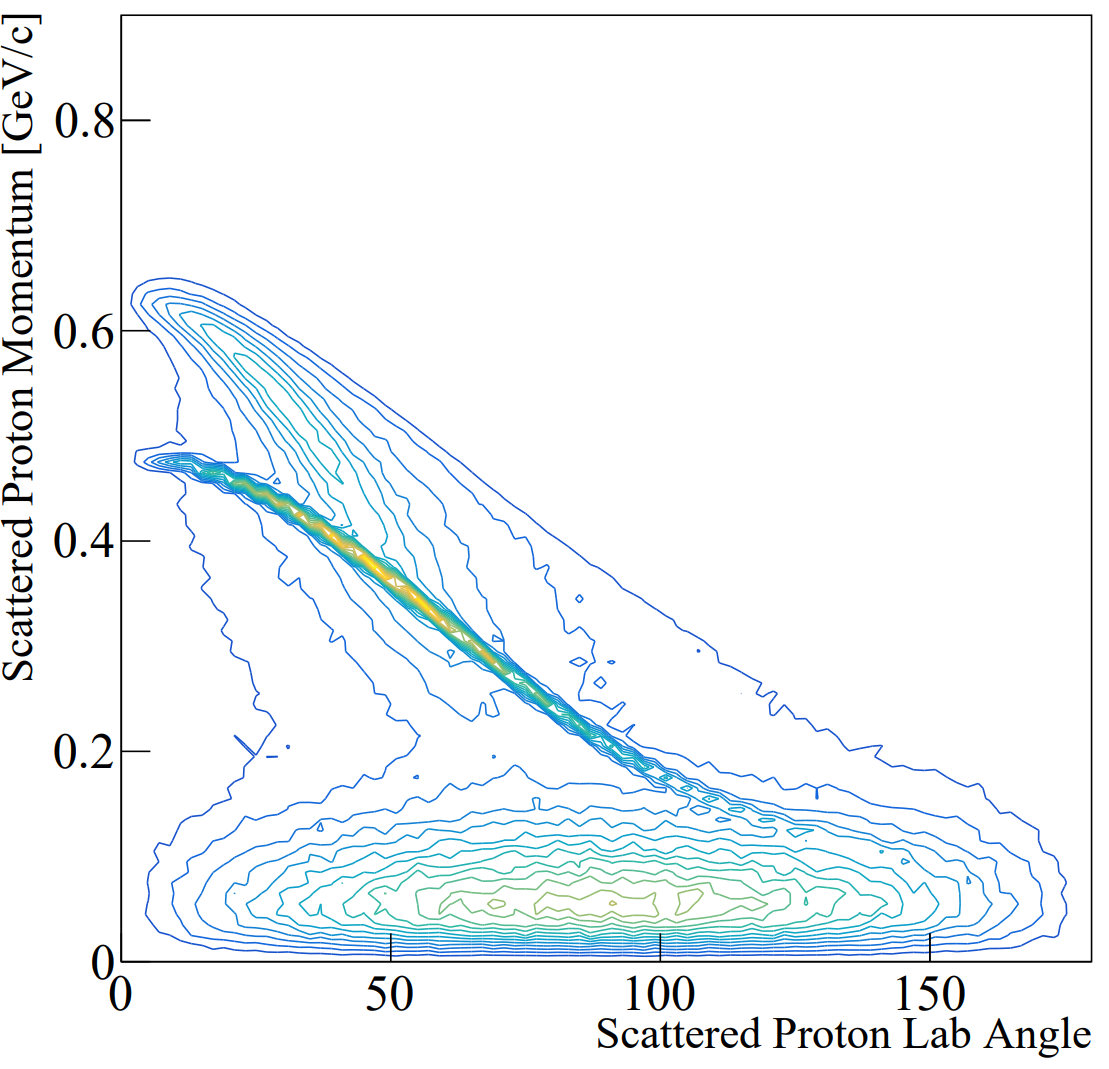} 
}}
}
\caption{(a) Momentum transferred to $\Theta^+$ in the $K^+d\to\Theta^+p$ and
$K^+p\to\Theta^+\pi^+$ reactions at lab scattering angles of $0^\circ$, $10^\circ$, and $20^\circ$; (b) a scatter plot of the momentum and lab scattering angle for the scattered protons in the
$K^+d\to\Theta^+p$ reaction. }
\label{fig:kine}
\end{figure}

The $K^+d\to\Theta^+p$ reaction can produce a recoiless $\Theta^+$ 
at a scattering angle of $0^\circ$, as illustrated in Fig. \ref{fig:kine}(a). This magic momentum 
is approximately 0.5 GeV$/c$, whichi is suitable for the $\Theta^+$ search experiment. This
ideal momentum allows the $K^0$ and $p$ to emerge in a back-to-back configuration. 
In contrast, the $K^+p\to\Theta^+\pi^+$ reaction has 
a minimum recoil momentum of about 0.6 GeV$/c$. A clear signal is expected from
this reaction due to the two-body kinematical correlation between the production angle and momentum, as shown in Fig. \ref{fig:kine}(b).

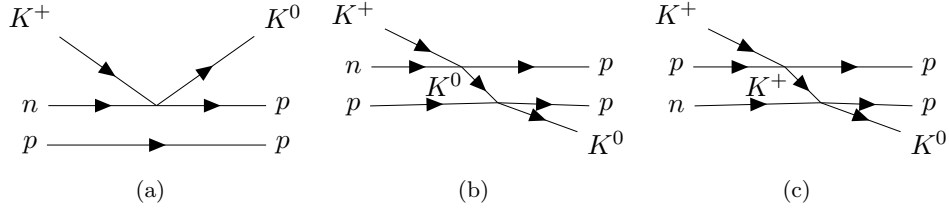
\begin{figure}[!hbtp]
\centering
\scalebox{0.95}{
\subfigure[]{
\begin{tikzpicture}
\begin{feynman}
\vertex at (0, 1) (i1) {\(K^+\)};
\vertex at (0,-0.25) (i2) {\(n\)};
\vertex at (0,-0.8) (i3) {\(p\)};
\vertex at (1.75, -0.25) (a);
\vertex at (3.5, 1) (f1) {\(K^0\)};
\vertex at (3.5,-0.25) (f2) {\(p\)};
\vertex at (3.5,-0.8) (f3) {\(p\)};
\diagram{(i1) -- [fermion] (a),
(i2) -- [fermion] (a),
(i3) -- [fermion] (f3),
(a) -- [fermion] (f1),
(a) -- [fermion] (f2),};
\end{feynman}
\end{tikzpicture}
}\hskip-0.1cm
\subfigure[]{
\begin{tikzpicture}
\begin{feynman}
\vertex at (0, 1) (i1) {\(K^+\)};
\vertex at (0, 0.25) (i2) {\(n\)};
\vertex at (0,-0.3) (i3) {\(p\)};
\vertex at (1.25, 0) (c) {\(K^0\)};
\vertex at (1.5,0.25) (a);
\vertex at (2.0,-0.25) (b);
\vertex at (3.5, -0.8) (f1) {\(K^0\)};
\vertex at (3.5,0.25) (f2) {\(p\)};
\vertex at (3.5,-0.3) (f3) {\(p\)};
\diagram{(i1) -- [fermion] (a),
(i2) -- [fermion] (a),
(i3) -- [fermion] (b),
(a) -- [fermion] (b),
(b) -- [fermion] (f1),
(b) -- [fermion] (f3),
(a) -- [fermion] (f2),};
\end{feynman}
\end{tikzpicture}
}\hskip-0.1cm
\subfigure[]{
\begin{tikzpicture}
\begin{feynman}
\vertex at (0, 1) (i1) {\(K^+\)};
\vertex at (0, 0.25) (i2) {\(p\)};
\vertex at (0,-0.3) (i3) {\(n\)};
\vertex at (1.25, 0) (c) {\(K^+\)};
\vertex at (1.5,0.25) (a);
\vertex at (2.0,-0.25) (b);
\vertex at (3.5, -0.8) (f1) {\(K^0\)};
\vertex at (3.5,0.25) (f2) {\(p\)};
\vertex at (3.5,-0.3) (f3) {\(p\)};
\diagram{(i1) -- [fermion] (a),
(i2) -- [fermion] (a),
(i3) -- [fermion] (b),
(a) -- [fermion] (b),
(b) -- [fermion] (f1),
(b) -- [fermion] (f3),
(a) -- [fermion] (f2),};
\end{feynman}
\end{tikzpicture}
}
}
\caption{The production processes of $\Theta^+$ in the $K^+d\to K^0pp$ reaction include; (a) a direct one-step process (impulse scattering), two-step processes involving (b) a charge-exchange reaction followed by elastic scattering, and (c) vice versa.
}
\label{fig:kd}
\end{figure}

A new experiment is being discussed to search for the $\Theta^+$ in the $K^+d\to K^0pp$ reaction 
at a momentum of 0.5 GeV$/c$ at J-PARC \cite{ahn}. This experiment was initially designed 
to exclusively measure the decay products $K^0p$ of the $\Theta^+$ along with a spectator proton.
The $\Theta^+$ can be produced via impulse scattering (Fig. \ref{fig:kd}(a)),
two-step processes involving a primary reaction $K^+n\to K^0p$ followed by a secondary 
reaction $K^0p\to K^0p$ (Fig. \ref{fig:kd}(b)) and the other with a proton first involved 
in the primary reaction (Fig. \ref{fig:kd}(c)).
\begin{figure}[htb]
\centerline{%
\includegraphics[width=0.475\textwidth]{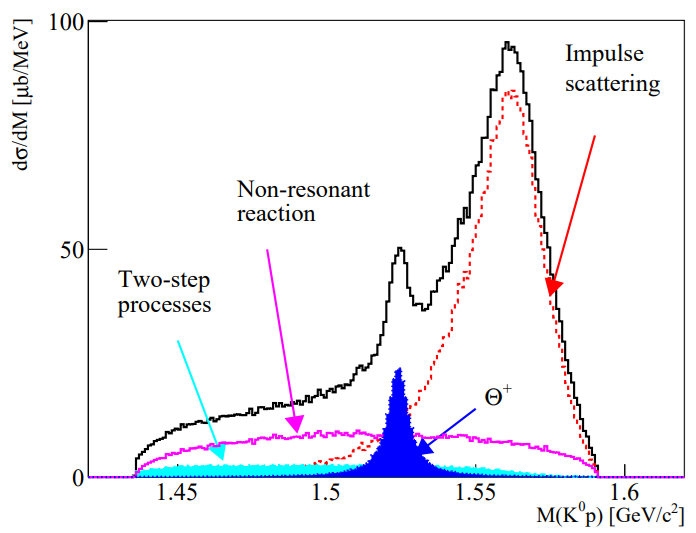} 
\hskip+0.2cm
\includegraphics[width=0.475\textwidth]{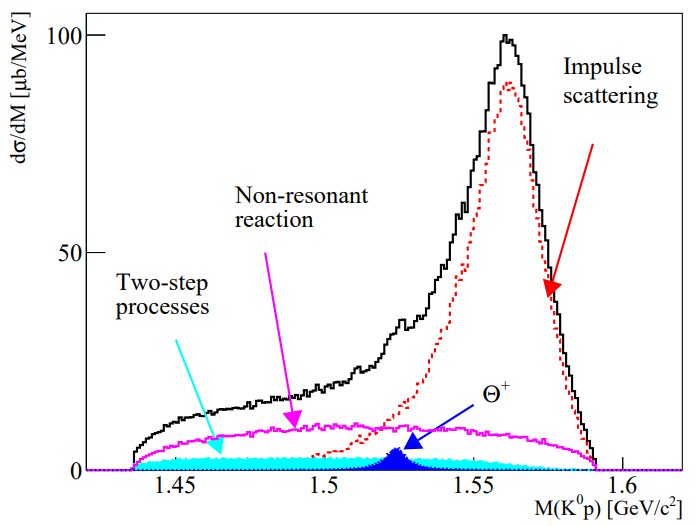} 
}
\caption{Expected $K^0p$ mass spectra for the $K^+d\to K^0pp$ reactions at 0.5 GeV$/c$, assuming the $\Theta^+$ production cross section is (a) 500 $\mu$b and (b) 100 $\mu$b, respectively. A 3 MeV mass resolution is taken into account. }
\label{fig:kpp}
\end{figure}

However, the use of an 80-mm thick cylindrical liquid deuterium target poses a challenge, 
as it limits the detection of spectator protons that emerge slowly from the target. 
Additionally, the intensity of the $K^+$ beam decreases significantly at low momenta. 
To address this issue, an alternative approach is being considered, which involves 
using a high momentum beam equipped with
a graphite degrader. This approach can increase statistics since higher momentum beams 
provide greater intensity \cite{test}. 

Nonetheless, this alternative means that the original goal of conducting a complete
kinematic analysis in the search for the $\Theta^+$ cannot be achieved 
because the exact beam momentum would remain unknown. In this case, the $K^0p$ mass resolution
will entirely depend on the momentum and angular resolutions in reconstructing the 
$K_S\to\pi^+\pi-$ decay and a fast proton. The expected $K^0p$ mass spectra are demonstrated
in Fig. \ref{fig:kpp}(a) and (b), assuming 3 MeV of the mass resolution for 
$\sigma_\Theta=500~\mu$b and 100 $\mu$b, respectively. The enhanced lineshape could only be
recognized when the $\Theta^+$ production cross section exceeds 100 $\mu$b. 

To effectively address this challenge, a reliable solution is to change 
the shape of the target cell from a cylinder to a long, thin slab. 
This adjustment would allow low-energy spectator protons 
to escape more easily from the target at large angles. Furthermore, 
the cross section measurement for the $K^+d\to K^0pp$ reaction is crucial, as it provides
important information about the center-of-mass energy for the $KN$ system,
particularly near the mass of the $\Theta^+$.

Additionally, it is essential to emphasize that a high-precision measurement of the 
$d(K^+,p)\Theta^+$ reaction has been proposed \cite{tanida}. 
This proposal can be improved by introducing a new approach 
to suppress background processes related to
elastic and charge exchange reactions using the HypTPC \cite{ichikawa}.   


There is a clear need in the near future for a new combined $I=0$ and $I=1$ partial-wave analysis
using all the new $K^+n$ data in combination with $I=1$ partial waves from $K^+p$ data.

\section{$\pi^-p$ reactions}

\begin{figure}[!hbtp]
\centering
\subfigure[]{
\begin{tikzpicture}
\begin{feynman}
\vertex at (0, 1) (i1) {\(\pi^-\)};
\vertex at (0,-1) (i2) {\(p\)};
\vertex at (1, 0) (a);
\vertex at (2,0) (b);
\vertex at (3, 1) (f1) {\(K^-\)};
\vertex at (3,-1) (f2) {\(\Theta^+\)};
\vertex at (1.5,0.5) () {\(N^\ast\)};
\diagram{(i1) -- [fermion] (a) -- [anti fermion] (i2),
(f1) -- [anti fermion] (b) -- [fermion] (f2),
(a) -- [fermion] (b),};
\end{feynman}
\end{tikzpicture}
}
\subfigure[]{
\begin{tikzpicture}
\begin{feynman}
\vertex at (0, 1) (i1) {\(\pi^-\)};
\vertex at (0,-1) (i2) {\(p\)};
\vertex at (1.5, 1) (a);
\vertex at (1.5,-1) (b);
\vertex at (3, 1) (f1) {\(K^-\)};
\vertex at (3,-1) (f2) {\(\Theta^+\)};
\vertex at (2.0,0.) () {\(K^\ast\)};
\diagram{(i1) -- [fermion] (a),
(i2) -- [fermion] (b),
(f1) -- [anti fermion] (a),
(b) -- [fermion] (f2),
(a) -- [fermion] (b),};
\end{feynman}
\end{tikzpicture}
}
\caption{(a) Feynman diagrams for $\Theta^+$ production in (a) the $s$ channel and (b) $t$ channel. 
}
\label{fig:pipdiagram}
\end{figure}
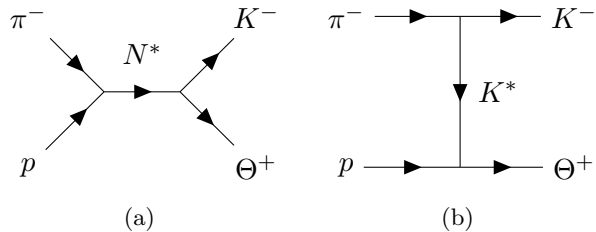

The $\pi^-p$ reaction can produce the $\Theta^+$ associated with $K^-$. Unlike the $\gamma p$ reaction, it can produce it with a proton target. Additionally, it can proceed only with a $K^\ast$ exchange in the $t$ channel, while the $\gamma p$ reaction can have both $K$- and 
$K^\ast$-exchange processes in the $t$ channel.   
The $\Theta^+$ production in $\pi^-p$ reactions can also occur through $N^\ast$ resonances 
in the $s$ channel above 
a threshold momentum of 1.7 GeV$/c$, as illustrated in Fig. \ref{fig:pipdiagram}(a) and (b).
However, the lack of observed $\Theta^+$ production in the $K^+p\to\Theta^+\pi^+$ reaction 
indicates that the $t$-channel process via the $K^\ast$ exchange is relatively small \cite{hyodo2}. 
Thus, it is likely that the $s$-channel contribution predominates 
in the $\pi^-p\to\Theta^+K^-$ reaction. 
  
\begin{figure}[htb]
\centerline{%
\stackinset{r}{1.6cm}{t}{3.2cm}{\small \textcolor{red}{$0^\circ$}}{
\stackinset{r}{0.7cm}{t}{2.2cm}{\small \textcolor{blue}{$10^\circ$}}{
\stackinset{r}{1.3cm}{t}{1.4cm}{\small \textcolor{darkcyan}{$20^\circ$}}{
\stackinset{l}{1.2cm}{t}{0.75cm}{\small \textcolor{black}{$\pi^-p\to\Theta^+K^-$}}{
\includegraphics[width=0.475\textwidth]{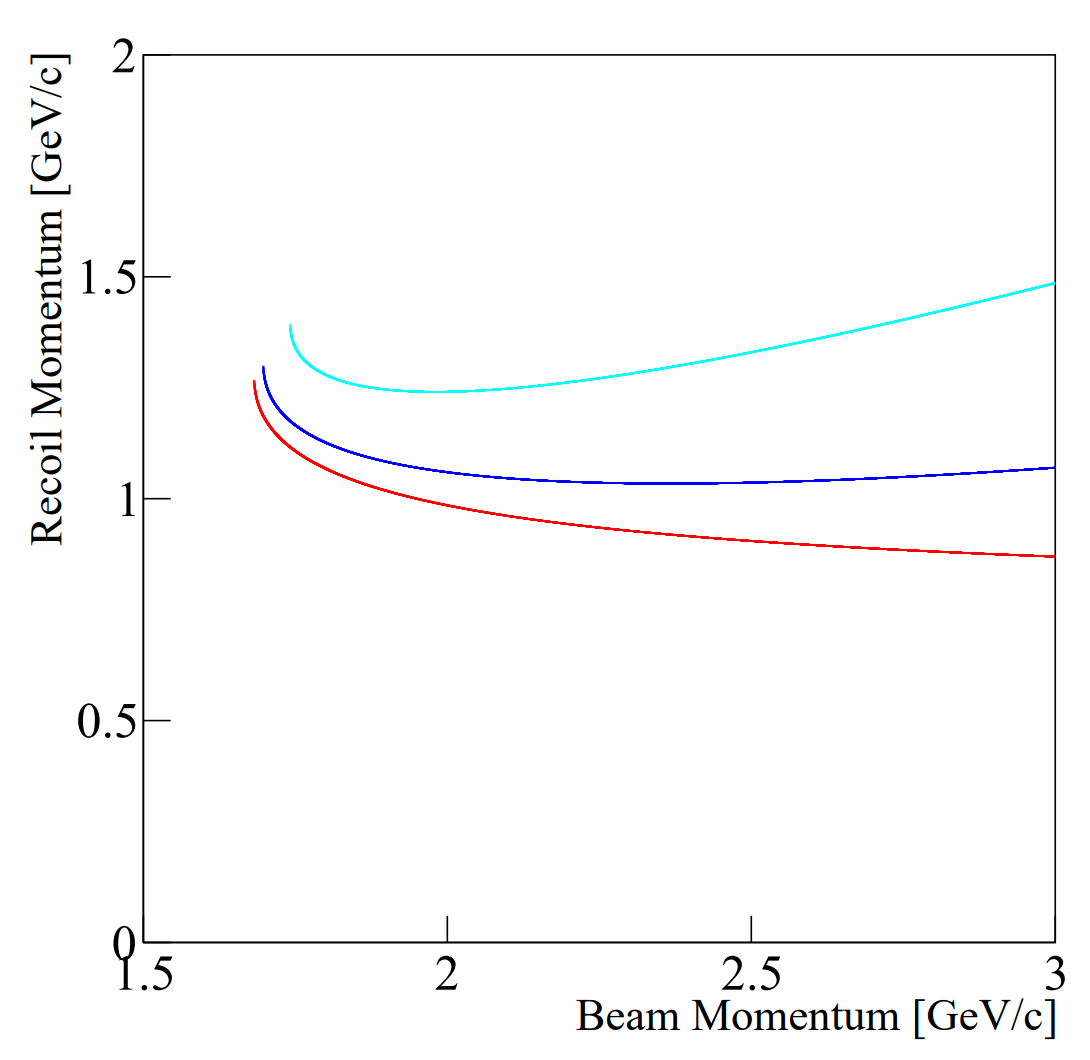} 
}}}}
\hskip+0.2cm
\stackinset{r}{1.6cm}{b}{1.7cm}{\small \textcolor{red}{$\phi n$}}{
\stackinset{r}{1.7cm}{t}{3.0cm}{\small \textcolor{blue}{$\Lambda(1520)K^+$}}{
\stackinset{r}{2.5cm}{b}{1.0cm}{\small \textcolor{darkcyan}{$\Theta^+$}}{
\stackinset{l}{1.0cm}{b}{1.5cm}{\small \textcolor{black}{non-resonant}}{
\stackinset{l}{1.6cm}{b}{1.1cm}{\small \textcolor{black}{$\downarrow$}}{
\includegraphics[width=0.475\textwidth]{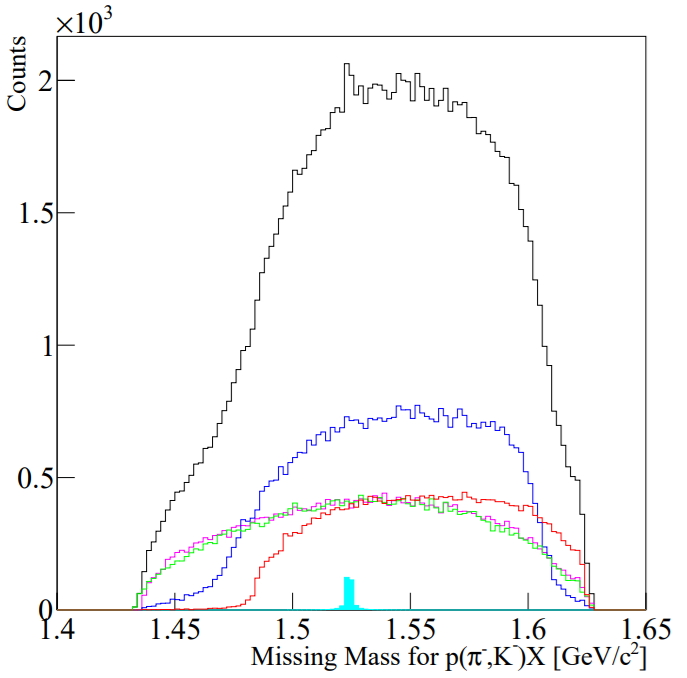} 
}}}}}
}
\caption{(a) Momentum transferred to the $\Theta^+$ in the $\pi^-p\to \Theta^+K^-$ reaction for different lab angles of $K^-$ as a function of beam momentum and 
(b) simulated missing-mass spectra for the $\pi^-p\to\Theta^+K^-$ reaction at 1.92 GeV$/c$.
}
\label{fig:pip}
\end{figure}

In the $\pi^-p\to\Theta^+K^-$ reaction, the $\Theta^+$ is produced with high momentum, 
as illustrated in Fig. \ref{fig:pip}(a). 
The J-PARC E19 experiment conducted a high-resolution and high-statistics measurement 
to search for the $\Theta^+$ in the $(\pi^-, K^-)$ reaction at 1.92 \cite{shirotori} and 
2.01 GeV$/c$ \cite{e19b}. The mass resolutions were $1.72$ and 2.13 MeV$/c^2$ (FWHM).
No peak structure was observed in the missing mass spectra at scattering angles 
ranging from $2^\circ$ to $15^\circ$ in the laboratory frame.
As a result, the upper limit on the forward production cross section 
for the possible $\Theta^+$ mass region was found to be less than
0.28 $\mu$b/sr at a 90\% confidence level.  

The $(\pi^-, K^-)$ reaction includes background processes such as $\phi$ and $\Lambda(1520)$
production reactions. Although the E19 accounted for these background contributions in the
missing-mass spectrum, there remains the potential for interference effects among the 
resonances. 
The simulated missing-mass spectra for the $\pi^-p\to\Theta^+K^-$ reaction at 1.92 GeV$/c$ are
shown in Fig. \ref{fig:pip}(b). 
This simulation assumes a $\Theta^+$ production cross section of $0.1~\mu$b 
with flat angular distributions for all processes.
It also considers the acceptance of the HypTPC, which is nearly $3\pi$. 
The $\Theta^+$ peak is represented as an incoherent sum of all amplitudes 
in Fig. \ref{fig:pip}(b). 
If the $\Theta^+$ production amplitude interferes destructively with other amplitudes, 
it could diminish the visibility of the $\Theta^+$ signal.  
Therefore, it is essential to measure all final state particles to investigate
the interference pattern in the Dalitz plot, particularly near the $\Theta^+$ mass.

\begin{figure}[!htb]
\centerline{%
\includegraphics[width=0.475\textwidth]{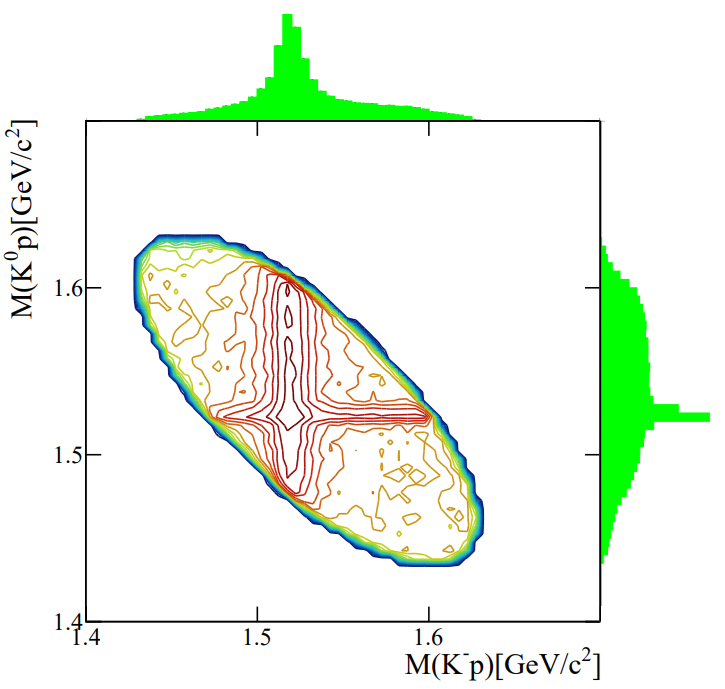} 
\hskip+0.2cm
\includegraphics[width=0.475\textwidth]{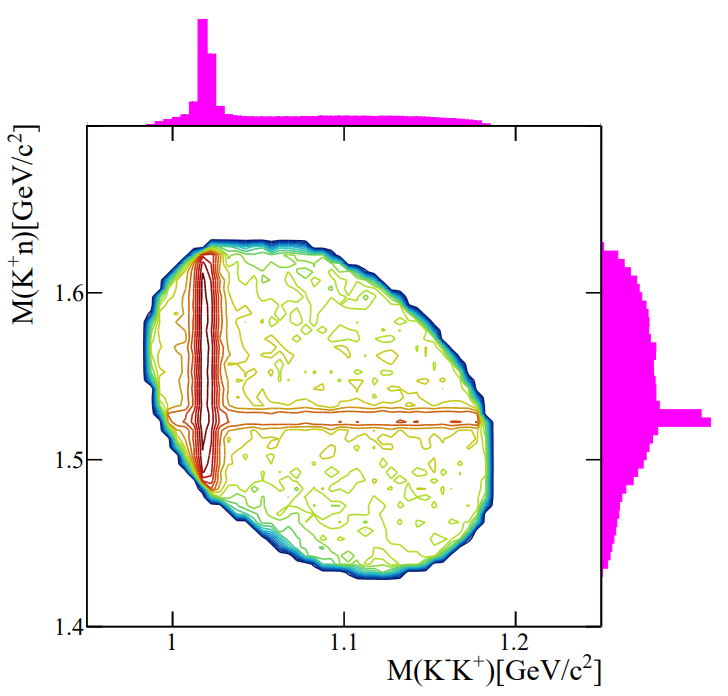} 
}
\caption{Simulated scatter plots for (a) $K^-p$ vs $K^0p$ mass distributions and 
(b) $K^-K^+$ vs $K^+n$ mass distributions for the $(\pi^-,K^-)$ reactions at 1.92 GeV$/c$. }
\label{fig:pip19}
\end{figure}

The two reactions $\pi^-p\to K^0K^-p$ and $\pi^-p\to K^+K^-n$ are
associated with the $\Theta^+$ production. 
The first reaction is linked to $\Lambda(1520)$, while the second is related to $\phi$ production.
Fig. \ref{fig:pip19}(a) and (b) show the simulated 
mass distributions for these two reactions at 1.92 GeV$/c$ are represented. 
This simulation assumes a $\Theta^+$ production cross section of $3.5~\mu$b, 
which highlights a distinct peaking structure. 

Both plots depict how the $\Theta^+$ band intersects with the bands for $\phi$ and $\Lambda(1520)$ 
resonances. This overlap necessitates the isolation of the background resonances 
by reconstructing all final state particles.
Notably, the crossing between the $\Lambda(1520)$ and $\Theta^+$  bands is broader than 
the crossing between the $\phi$ and $\Theta^+$ bands, which could lead to a more pronouced interference effect.
  
The $\Theta^+$ band starts to diverge from the $\phi$ and $\Lambda(1520)$ bands at 2.2 GeV$/c$, 
where no interference effects are observed, as shown in Fig. \ref{fig:pip19}(a) and (b). 
There may be other resonances in the high mass region,
such as $\Sigma(1670)$ and $f_2(1270)$, but these are weakly coupled to 
$K^-p$ and $K^+K^-$ channels. The measurement of the $\pi^-p\to\Theta^+K^-$ reaction
will be available in the $\pi 20$ beam line at J-PARC.

\begin{figure}[!htb]
\centerline{%
\includegraphics[width=0.475\textwidth]{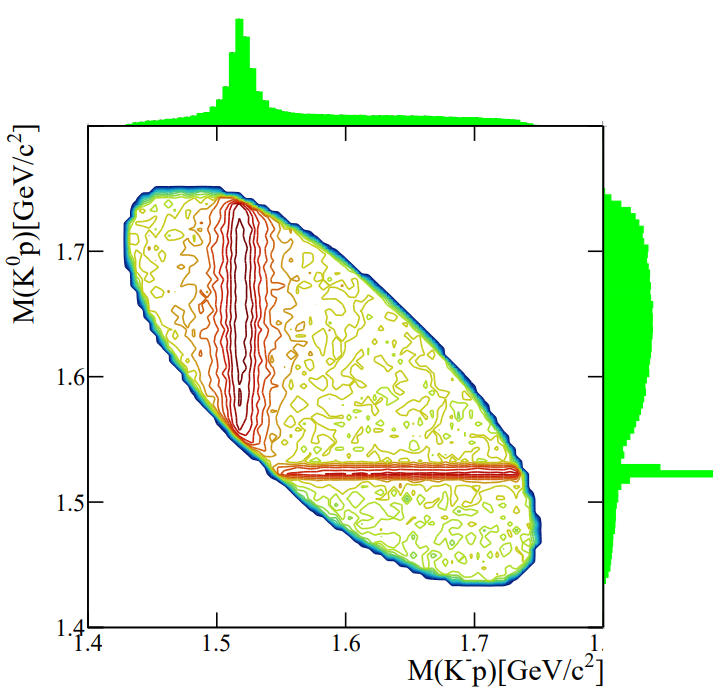} 
\hskip+0.2cm
\includegraphics[width=0.475\textwidth]{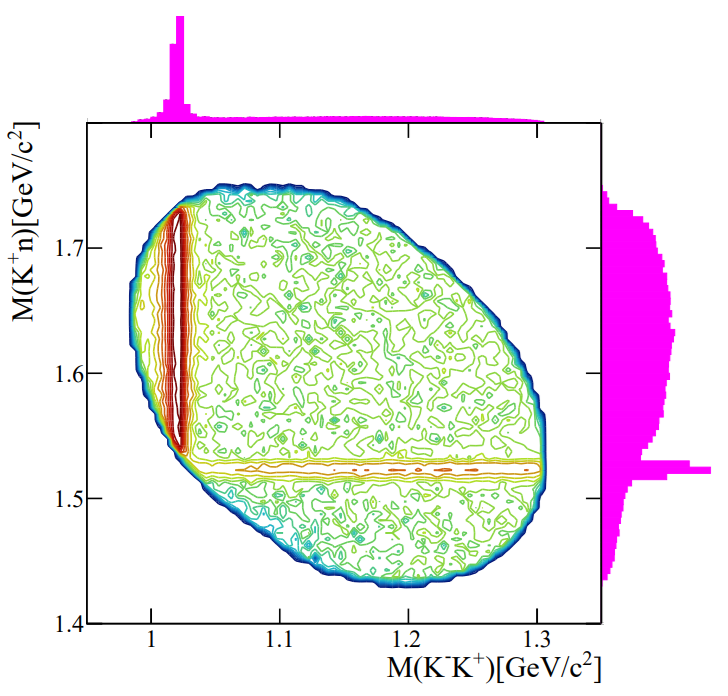} 
}
\caption{Simulated scatter plots for (a) $K^-p$ vs $K^0p$ mass distributions and 
(b) $K^-K^+$ vs $K^+n$ mass distributions for the $(\pi^-,K^-)$ reactions at 2.2 GeV$/c$. }
\label{fig:pip22}
\end{figure}

\section{Hidden-Strangeness Pentaquarks}

Hidden-strangeness pentaquarks are considered siblings to $P_{c\bar{c}}$ states that were observed in the 
decays of $\Lambda_b\to J/\psi pK^-$ at LHCb. A new antidecuplet, formed by combining 
vector mesons with baryon octet members, includes an isospin doublet of $P_{s\bar{s}}$ states 
and a singlet $P_{\overline{s}}$ state.   
If we assume a similar mass splitting between $P_{\overline{s}}$ 
and $P_{s\bar{s}}$ states, 
we find that $P_{\overline{s}}$ has a mass of 1.91 GeV.
This state can be searched for in photoproduction and hadron-induced reactions. 

The search for the $P_{s\bar{s}}^+$ state was conducted in the Cabbibo-suppressed decays of 
$\Lambda_c^+\to \phi p\pi^0$ at Belle\cite{belle} and BESIII \cite{bes3}. 
Due to limited statistics, the results provided only an upper limit on the combined branching ratio
of $\mathcal{B}(\Lambda_c^+\to P_{s\bar{s}}^+\pi^0)\cdot\mathcal{B}(P_{s\bar{s}}^+\to\phi p)$ to be 
$8.3\times 10^{-5}$ at 90\% confidence level. 
Although the $\Lambda_c^+$ can decay to $\phi p\pi^0$ decay 
with a sufficient phase space of approximately 100 MeV, this decay is significantly suppressed. 

This suppression can be attributed to the dominance of the triangular singularity diagram in the
decay $\Lambda_c^+\to\phi p\pi^0$ via $\Sigma^{\ast +}K^{\ast 0}$ \cite{he}. 
The $\Sigma^{\ast +}$ decays into $\Sigma^+\pi^0$, 
and the $\Sigma^+$ then interacts with $K^{\ast 0}$ to produce $\phi$ and $p$.  
In this scenario, the available phase space for the decay $\Lambda_c^+\to\Sigma^{\ast +}K^{\ast 0}$ 
is limited to just a few MeV, which results in a highly suppressed branching ratio.

The $\phi p$ bump structure could be interpreted as a $\Sigma K^\ast$ molecular state 
($J^P=3/2^-$). This interpretation explains the $\Sigma^+K^{\ast 0}\to \phi p$ reaction 
in the decay $\Lambda_c^+\to\phi p\pi^0$, 
as well as the near-threshold reaction $\gamma p\to \phi p$ \cite{xie}. 
In addition, the measured parity spin asymmetry in the 
$\gamma p\to K^{\ast 0}\Sigma^+$ reaction supports the dominance of 
a natural-parity exchange process, which shares characteristics with Pomeron exchange \cite{hwang}.  

In the $\gamma p\to\phi p$ reaction, the $t$-channel process dominates, 
while in the $\pi^-p\to\phi n$ reaction, the $s$ channel are more significant. 
The $\pi^-p$ reaction specifically investigates the $P_{s\bar{s}}^0$, which is an isospin partner of
$P_{s\bar{s}}^+$.  
A measument of  the $\pi^-p\to\phi n$ reaction has been proposed
using a secondary $\pi^-$ beam with momentum ranging from $1.6$ to $2.4$ GeV$/c$, 
delivered by the $\pi$20 beam line \cite{p95}. In this reaction, only the production processes of $\phi$ and $\Sigma(1775)$ 
can contribute to the final state $K^+K^-n$. 
This new proposal, referred to as J-PARC P95, also aims to measure the production of 
$K^\ast\Lambda$, $K\Lambda^\ast$, $K^\ast\Sigma$ and $K\Sigma^\ast$. 
These processes are likely to have a strong coupling to the $\phi p$ channel.

In addition, the $P_{s\bar{s}}$ production reactions using hadron beams can also be explored. 
The $\pi^-p\to \pi^-p\phi$ reaction is available above a threshold momentum of 1.86 GeV$/c$.
In the $K^+p\to K^+p\phi$ reaction, the only
background comes from $K^\ast(1680)$, which is just below $K^\ast\Sigma$ threshold \cite{sinam}.
Therefore, this reaction is one of the most promissing reactions 
to investigate the existence of the $P_{s\bar{s}}$.

The $P_{\bar{s}}$ is located at the top corner of the antidecuplet triangle, 
sharing the same location as the $\Theta^+$ in the antidecuplet multiplet formed by 
the combination of pseudoscalar mesons and baryon octet members. 
An observation of the $f_1(1285)$ meson has been reported in the $\pi^-K^+\bar{K}^0$ 
channel in the $\gamma p\to pK^+\pi^-\bar{K}^0$ reactions 
using the CLAS detector at JLab \cite{clas2}. 
Within the same dataset, there may be a hint of the $P_{\bar{s}}$ signal, if it indeed exists.

This state could also be investigated in the decay channels involving $K^\ast N$ through
the $\gamma p\to P_{\bar{s}}(1910)\bar{K}^0$ reactions, 
which occurs above a threshold energy of 2.65 GeV,
and $\pi^-p\to P_{\bar{s}}(1910)^+K^-$, which takes place above a threshold momentum 
of 2.6 GeV$/c$.

\section{Summary}

The $\Theta^+$ search should continue until its existence is definitely ruled out
in hadron-induced reactions, such as $K^+p\to\Theta^+\pi^+$, $K^+d\to K^0pp$, 
and $\pi^-p\to\Theta^+K^-$. 
New measurements need be designed to reconstruct all final states at energies 
where the $\Theta^+$ band does not overlap with background resonance bands. 
A direct formation of $\Theta^+$ in the $K^+n\to K^0p$ reaction using a deuterium target 
would provide a straightforward way to confirm the existence of the $\Theta^+$; 
however, the final state interaction between $pp$ 
and the measurement of spectator protons present significant challenges. 

To provide conclusive evidence for either the existence or exclusion of the $\Theta^+$, 
a new experimental approach must achieve a sensitivity that is 
two orders of magnitude higher than previous $\Theta^+$ searches. 
This unprecedented experiment
will put an end to the $\Theta^+$ controversy.

\end{document}